# On the Contents and Utility of IoT Cybersecurity Guidelines

JESSE CHEN, University of Arizona, USA
DHARUN ANANDAYUVARAJ, Purdue University, USA
JAMES C. DAVIS, Purdue University, USA
SAZZADUR RAHAMAN, University of Arizona, USA

Cybersecurity concerns of Internet of Things (IoT) devices and infrastructure are growing each year. In response, organizations worldwide have published IoT security guidelines to protect their citizens and customers by providing recommendations on the development and operation of IoT systems. While these guidelines are being adopted, *e.g.* by US federal contractors, their content and merits have not been critically examined. Specifically, we do not know what topics and recommendations they cover and their effectiveness at preventing real-world IoT failures.

In this paper, we address these gaps through a qualitative study of guidelines. We collect 142 IoT cybersecurity guidelines and sample them for recommendations until reaching saturation at 25 guidelines. From the resulting 958 unique recommendations, we iteratively develop a hierarchical taxonomy following grounded theory coding principles and study the guidelines' comprehensiveness. In addition, we evaluate the actionability and specificity of each recommendation and match recommendations to CVEs and security failures in the news they can prevent. We report that: (1) Each guideline has gaps in its topic coverage and comprehensiveness; (2) 87.2% recommendations are actionable and 38.7% recommendations can prevent specific threats; and (3) although the union of the guidelines mitigates all 17 of the failures from our news stories corpus, 21% of the CVEs evade the guidelines. In summary, we report shortcomings in each guideline's depth and breadth, but as a whole they address major security issues.

## 1 INTRODUCTION

Engineering guidelines constrain the engineering process to improve the quality of the resulting system [70]. One class of software that has been the particular focus of engineering guidelines is Internet of Things (IoT) systems, also known as cyber-physical systems [76, 78, 114]. IoT systems use software to handle sensitive data and control sensitive resources, thus their security and privacy assurances are paramount. When IoT systems fail to meet these requirements, attackers can leverage

Authors' addresses: Jesse Chen, University of Arizona, Tucson, USA, jessechen@arizona.edu; Dharun Anandayuvaraj, Purdue University, West Lafayette, USA, dananday@purdue.edu; James C. Davis, Purdue University, West Lafayette, USA, davisjam@purdue.edu; Sazzadur Rahaman, University of Arizona, Tucson, USA, sazz@cs.arizona.edu.

the IoT systems for serious harm [106], from crashing cars [6, 13] to botnets [20]. To reduce the risks of IoT systems, following the path of older software and systems standards such as ISO 9001 (quality) and IEC 61508 (safety-critical systems), cybersecurity guidelines for IoT engineering have been published by governments, non-profits, and industry leaders. For example, the USA's IoT Cybersecurity Improvement Act of 2020 [14] mandated that federal contractors follow IoT security guidelines from the US National Institute of Standards and Technology (NIST) [90, 91, 118]. IoT manufacturers and service providers are applying these guidelines [35, 38, 108]. Adopting and demonstrating compliance with these guidelines is costly [35].

Despite widespread industry adoption, there has been little independent evaluation of these IoT security guidelines. We know of only three works: Momenzadeh *et al.* studied 6 guidelines and assessed their benefits in two IoT market hubs [86], Barrera *et al.* assessed one guideline [116] for actionability [34], and Bellman & van Oorschot compared two of these guidelines [36]. Two gaps in our knowledge are: (1) *What security aspects are covered by these guidelines, both individually and in aggregate?* and (2) *Can the guidelines mitigate real IoT security failures?* Guideline adopters and guideline publishers would benefit from this knowledge. For example, adopters can learn which guidelines might benefit them, and publishers can improve their guidelines.

To address this gap, we systematically collected and analyzed a set of IoT cybersecurity guidelines using a qualitative approach [51]. Specifically, we first collected 142 general IoT Security guidelines, as they offer the basis for industry-specific guidelines and dictate baseline security. We sampled 25 of these guidelines (∼900 pages) and manually extracted 958 unique (1,514 total) security recommendations (§3). We then organized these recommendations into a hierarchical taxonomy, proceeding with iterative sampling and taxonomization until saturation [113]. We followed a grounded theory framework [44, 113] to guide this process. This choice was guided by the inductive nature of the study and the unstructured nature of the data. To evaluate the recommendations in the guidelines, we quantitatively assessed each guideline's comprehensiveness by comparing them to the entire collection of guidelines (§4.1) and the topic coverage of the taxonomy (§4.2). To evaluate the usefulness of the guidelines, we evaluated each recommendation's actionability and specificity (§5.1), both ourselves and via a user study with novice (student) software engineers. Finally, with a retrospective study (§5.2), we investigated the recommendations' potential to address real-world IoT security failures, as documented in CVEs and by the news media.

We report shortcomings in each guideline's individual depth and breadth and insights into their ability to prevent security issues as a whole. Regarding recommendations, even the most verbose individual guideline was not comprehensive. In terms of usefulness, many of the recommendations are actionable (836/958, 87.2%), while less than half were objectively specific (371/958, 38.7%) to security concerns. The unified taxonomy resulted from our study could only mitigate 79% (77/97) of the studied CVEs but all 17 studied news articles. This informs stakeholders that following general guidelines alone is insufficient while also showing opportunities for many publishers to improve their guideline coverage. **In summary, our contributions are**:

- **Unified IoT Security Guidelines**: We systematically sampled and analyzed 25 guidelines to extract 1,514 IoT security recommendations (958 unique) and organize them hierarchically.
- **Recommendations**: First, we compared the taxonomy of recommendations to the individual guidelines to determine the comprehensiveness of the guidelines individually. Then, we analyze the collective topic coverage by guideline publisher type. Our results show that even the guidelines with the most recommendations suffer from a lack of depth.
- **Usefulness**: First, we measured the number of recommendations that were specific and actionable. Then, we applied the guidelines to real-world IoT security failures. Our results show that 79% of the CVEs and 100% of the news story failures can be prevented.

## 2 BACKGROUND, PRIOR WORK, AND OUR RESEARCH
### 2.1 IoT Security Guidelines
Internet of Things (IoT) systems are heterogeneous distributed systems. IoT engineering teams have limited resources and do not always prioritize non-functional requirements [35, 103]. Thus, security is an afterthought in many IoT systems [28, 55], or incorporated on demand [62]. Studies describe many issues in IoT development that may lead to insecure IoT systems [32, 79].

To enable baseline security, many organizations have published IoT security guidelines. These organizations include government bodies such as NIST [89], for-profit companies such as Microsoft [25], non-profit trade groups such as the Cloud Security Alliance (CSA) [46, 47], and professional organizations such as the Institute of Electrical and Electronics Engineers (IEEE) [68].

We define an **IoT cybersecurity guideline** as a set of "best practice" **recommendations** to secure IoT production and use, *i.e.* the security capabilities to incorporate into the product, and the engineering and operational processes for security. These guidelines can be **general** or **domain-specific** (*e.g.* healthcare). Domain-specific guidelines usually extend general guidelines. These general guidelines generally define IoT as a device or system of devices that are connected via some network (*e.g.* internet, Bluetooth). Thus, their threat models are similar, consisting of network threats (*e.g.* man-in-the-middle attacks) [68] and device threats (*e.g.* unauthorized access) [116].

Several governments are moving towards mandating these guidelines at different levels. In the United States, there has been state-level legislation (*e.g.* California [7], Oregon [11], Virginia [18]), as well as national-level legislation [14] to regulate IoT security. Other governments have also enacted legislation to regulate IoT security: UK [22], Singapore [1], and the European Union [19]. Individual companies are also adopting certain guidelines — for example, in 2023, GE Gas Power adopted the NIST Framework for Improving Critical Infrastructure Cybersecurity [21].

### 2.2 Prior Research on IoT Security Guidelines
Prior research on IoT security guidelines has primarily focused on developing new guidelines [26, 27, 45, 97]. While these efforts are valuable, we believe that IoT engineers are more likely to rely on guidelines published by governments and industry leaders. Despite their practical importance, these formal IoT Security Guidelines have not received much research attention. The UK government extracted and mapped recommendations from IoT security guidelines onto their 13 code of practice guideline [15, 116], which misses many recommendations not falling within the 13 codes (§7). Barrera *et al.* conducted an actionability study of the extracted recommendations from the UK government's guideline [34]. Bellman *et al.* used Barrera *et al.*'s method to study the actionability of two other IoT security guidelines [36]. Momenzadeh *et al.* analyzed the security of two IoT marketplaces, reporting that the security violations could have been mitigated through the use of 6 distinct IoT cybersecurity guidelines [86].

### 2.3 Research Questions and Scope
In summary, although IoT cybersecurity guidelines are being widely adopted, prior works do not fully elucidate their content (topics covered and level of detail) nor utility (mitigation of real-world IoT security failures). Our research addresses this gap under two themes:

*Theme I: Guideline Contents.*
- ***RQ1: (§4.1)*** *How comprehensive (breadth and depth coverage) are individual IoT guidelines?*
- ***RQ2: (§4.2)*** *What topics are covered collectively? How does coverage vary by publisher type?*

*Theme II: Guideline Utility.*
- **RQ3 (§5.1):** *To what extent are the recommendations concrete enough to be actionable and specific enough to address distinct security threats, vulnerabilities, or attack surfaces?*
- **RQ4 (§5.2):** *To what extent could the guidelines have prevented real-world security problems?*

**Study Scope.** This study aims to understand the quality of the contents and the utility of IoT guidelines. There are many such guidelines and studying all of them comprehensively is beyond the scope of a single study. As a starting point, we focus specifically on general IoT guidelines. These guidelines are important to study because they dictate baseline security across all domains, and provide the foundation for domain-specific guidelines [46, 47, 89]. For instance, NIST-2021 says, "*IoT devices in-scope for this publication can function on their own, although they may be dependent on other specific devices (e.g., an IoT hub) or systems (e.g., a cloud) for some functionality*" [89]. This indicates that the guideline is generally applicable to all IoT devices, irrespective of their domains.

This general-specific relationship imitates other domains, such as: (1) in the Payment Card Industry, the Data Security Standard [10] is a general guideline that many sub-guidelines extend [5]; and (2) in safety-critical systems, the International Electrotechnical Commission's standard IEC-61508 [4] is used as a baseline, and different regulators provide domain-specific guidance [3].

## 3 HIERARCHICAL TAXONOMY OF RECOMMENDATIONS

**Overview.** Figure 1 gives an overview of our methodology for taxonomy construction. We collect guidelines via Internet searches (§3.1). We sample guidelines to iteratively construct the taxonomy until reaching saturation (§3.2). We then validate the taxonomy to increase confidence (§3.3). In §3.4 we present the taxonomy, and in subsequent sections we analyze it based on our research questions.

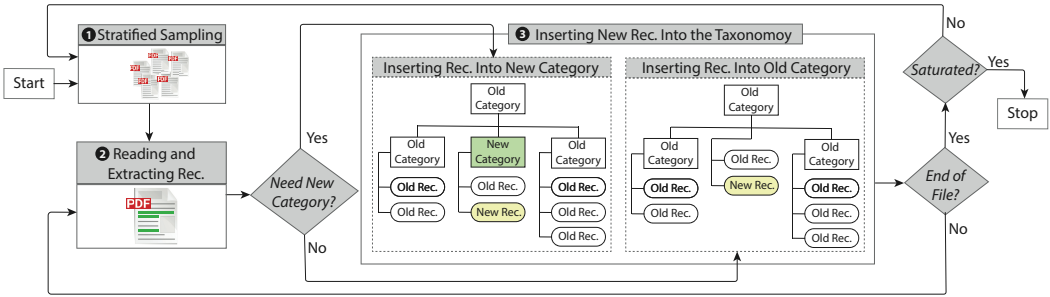

Fig. 1. Illustration of the taxonomy creation methodology flowchart. Part (1) is discussed in §3.2.2, part (2) in §3.2.3, and part (3) in §3.2.4; they represent an example subtree of the taxonomy. We repeat this entire process until we reach saturation (§3.2.6).

### 3.1 Guideline Collection

In this section, we describe our guideline collection methodology and results.

**Methodology.** Our goal was to collect a representative sample of general IoT security guidelines. Because there is no widely accepted source of IoT security guidelines, we opted to use two Internet search engines,[1] and a range of queries.[2] After each search, we examined the first five pages of results from each search engine (∼50 results per engine) for IoT security guidelines.

---
[1]We considered enumerating guideline-providing organizations such as professional societies (*e.g.* IEEE), governments (*e.g.* NIST), and companies (*e.g.* Intel), but this would bias results to organizations known to us [31].
[2]Search engines: Bing, Google. Filetype: PDF. Queries: "IoT security guidelines", "IoT security guidance", "IoT security guide", "IoT security recommendations", "IoT security best practices", "IoT security policies", and "IoT security".

Our exclusion criteria were non-English documents and documents that were unequivocally not guidelines (*e.g.* marketing material). Our inclusion criterion is that the document must provide relevant recommendations, *i.e.* descriptions of IoT device/system capabilities or organizational processes to promote security (§3.2.3). For each guideline, we visit the publisher's website to identify and collect only the newest version of the guideline. Like prior work, we acknowledge possible bias from the English-language restriction [98, 121].

**Results.** We collected 142 guidelines, with 63 from industrial for-profits (*e.g.* Microsoft, Amazon), 37 from governments (*e.g.* NIST, ENISA), 37 from industrial non-profits (*e.g.* Cloud Security Alliance, IoT Security Foundation), 3 from community non-profits (*e.g.* Open Web Application Security Project), and 1 from the professional organization IEEE. Our methodology resulted in guidelines from a wide variety of publishers from North America, Europe, Australia, and Asia.

### 3.2 Taxonomy Creation Methodology

After collecting the guidelines, we extracted and categorized the recommendations to unify them under a hierarchical taxonomy. We first conduct a pilot study (§3.2.1) to refine the subsequent methods. We then stratified-sample guidelines (§3.2.2) until saturation (§3.2.6). For each guideline, we extract recommendations (§3.2.3) and add them to the taxonomy (§3.2.4). In §3.2.5 we discuss our controls for subjective bias.

*3.2.1 Pilot Study.* Qualitative analysts should maintain an open mind [44] and should not inject personal assumptions, biases, or motivations [112]. With this in mind, we conducted a pilot study [51] by reviewing five randomly chosen guidelines to explore the existing landscape. Our observations led us to create two top-level categories for recommendations: (1) *Device Capabilities and Requirements* and (2) *Organizational Processes*. In the device capability category, recommendations prescribe a capability, feature, or configuration of the IoT device: *e.g.* "*Build devices to be compatible with lightweight encryption and security techniques*" [52]. On the other hand, in the process category, recommendations prescribe a process that should be followed by an entity to maintain secure development or operation: *e.g.* "*Train employees in good privacy and security practices*" [52]. Beyond this, from the pilot study, we confirmed that the documents did meet our definition (*i.e.* they contained IoT cybersecurity guidelines). We also noticed that these guidelines have no consistent file structure or format and use various terms for the same concepts, which makes automated text extraction analysis challenging [77, 125, 127].

*3.2.2 Why Sample Guidelines?* Manually reviewing all 142 guidelines (3804 pages), extracting recommendations, and constructing a taxonomy manually is cost-intensive. In qualitative studies of this complexity and scale, rigor must always be balanced against cost. To achieve this balance, we used a saturation-driven (details in §3.2.6) stratified sampling method [41, 51, 104, 112] to create a smaller representative subset of our guideline corpus. Saturation-driven sampling is often used in qualitative studies across numerous disciplines to build taxonomies, including software engineering [102]. We used publisher type (*e.g.* government vs. industrial) as the stratum for sampling, as they may have varying levels of concern in terms of security — *e.g.* governments may prioritize citizen rights, while companies may weigh profits. We determined the publisher type by looking on the publisher website's "about us" page. The sampled 25 guidelines are listed in the supplementary materials along with their abbreviated codenames.[3]

*3.2.3 Extracting Recommendations.* In qualitative data analysis, open coding begins with a text (*e.g.* interview transcript) and proceeds via memoization and/or perusing the transcript for keywords or phrases [85]. In our study, we studied the guidelines for recommendations in a sentence-by-sentence analysis, conducting a time-consuming but detailed investigation of the data [113]. As such, given a guideline document, we read it and identify phrases(s) or sentence(s) that give a security-related suggestion to an ecosystem entity. Recommendations can vary from adding a feature to an IoT device to following a process while building or using it. Definitions, descriptions, or explanations of terms, processes, and technologies are excluded by default since they are not actionable. In the following quoted text, the first sentence is not a recommendation, while the second is: "*Depending on the protocol and on available computing resources, a device may be more or less able to use strong encryption. Manufacturers should examine their situation on a case-by-case basis and use the strongest encryption possible, preferably IPsec and/or TLS/SSL*" [68]. Once a recommendation is identified, we record it in a spreadsheet. If a sentence contains more than one recommendation, *e.g.* separated with the word "*and*", we split them into multiple recommendations.

---

[3]Codenames use the format "[publisher name]-[year of publishing]". For example, a Microsoft guideline released in 2017 [25] has the codename "MSFT-2017". If the year of release is unknown, we use "20XX".

*3.2.4 Categorizing Recommendations and Creating Categories.* Our category creation and recommendation labeling process combines open and axial coding [113]. In open coding, similar concepts are grouped into distinct categories [113]. These categories are linked, and subcategories are created in axial coding [113]. To categorize each recommendation, we iteratively choose the most representative category at each taxonomy level (open coding). After categorizing a recommendation, we revisit the siblings of the category to see if any of them can be grouped into a more fine-grained sub-category, thereby creating new categories in the taxonomy. New categories are immediately created as subcategories of existing categories in the taxonomy (axial coding).

As might be expected, different guidelines sometimes contain equivalent recommendations. Each recommendation node in our taxonomy is a set of equivalent recommendations ("duplicates"). Here is an example pair: *"[Secure] device communications with secure technologies and protocols (e.g. TLS)"* [123] and *"Ensure that communication security is provided using state-of-the-art standardized security protocols, such as TLS for encryption"* [52]. Both recommend using standard protocols instead of proprietary ones, with TLS as an example.

*3.2.5 Reducing Bias via Weekly Collaboration.* Qualitative research necessitates the human researcher to be the data collection and coding instrument, with bias due to their life experiences and background [40]. To mitigate such biases, we leveraged collaborative qualitative data analysis [100]. During this process, the primary analyst consulted two other analysts on any confusion or questions regarding coding categories, recommendations, or recommendation placement in the taxonomy. In addition, these two other analysts also reviewed the developing taxonomy and provided feedback in case of any inconsistent categorizations.[4] This analyst team discussed all issues until reaching unanimous agreement. During the early stages of the weekly collaboration, which includes the pilot study, the discussions were focused on developing the first two levels of the taxonomy. Later on, the discussions shifted towards developing the deeper levels of the tree, categorizing recommendations, and optimizing the structure of categories in the taxonomy. These collaboration meetings took place at least weekly over 9 months (4/2022 to 12/2022).

*3.2.6 Saturation as a Stopping Point.* Saturation is a principle often used in grounded theory research to determine the stopping criteria [60]. The goal is to reliably sample the input space so that the most relevant concepts are covered, and new data is unlikely to contribute to new insights [60, 101]. Reaching saturation with strict criteria strengthens content validity [58].

Like other studies in software engineering research [66, 82, 84, 102, 121], we use this to determine the stopping criteria for guideline reviewing and recommendation collection (*i.e.* collecting more data). We assume that our recommendation corpus is saturated when all of the following requirements are fulfilled:

(1) The guideline, *G*, does not increase the unique recommendation count in the taxonomy by more than 5%. In other words, minimal new information is being added by reviewing more guidelines. To estimate a good stopping point, we looked at roughly how many new recommendations emerged in the guidelines from the pilot study. We anticipate that 5% is reasonably low, at which saturation is still achievable.

(2) The number of recommendations in *G* is larger than the median number of recommendations extracted from the previously read guidelines. This criterion prevents false saturation from guidelines with few recommendations.

(3) The constraints above must hold for three guidelines consecutively, as recommended (analogously) in [58]. This requirement gives another layer of protection against false saturation.

---

[4]To facilitate review and discussion, we created a web-based visualization tool for the taxonomy (see supplement).

**Validation of Saturation Criteria.** Saturation criteria are challenging to design [58], so there is a risk of missing categories if the remaining guidelines are substantially different in content. Therefore, to validate our saturation criteria, we reviewed the remaining 117 guidelines for new topics (*i.e.* potential new categories). In the validation process, a lower number of new topics (concepts) indicates a higher level of saturation, and vice versa. To identify topics in a guideline, we first refer to section titles. We review the body text if the section titles do not reflect the topics in the taxonomy, or if more information is needed to determine the topics covered. For each topic, we traverse the taxonomy to find a matching category. For example, most guidelines discuss network security, a category covered in our taxonomy. If we cannot find a match for a given topic, then we list it as a new topic. For instance, while we have a category for network security, we do not have any for telecommunications security (sim card, voice, and SMS), indicating that it is a new topic. This methodology offers a low-cost *estimate* of actual recommendation saturation. Directly validating recommendation saturation would require a detailed review of all guidelines, eliminating any benefit from using saturation as a stopping point.

Of the 100 unique topics (517 total) discovered, we found 7 new topics in the remaining 117 guidelines. The new topics are: telecommunication security, social engineering, environmental sustainability, designing for minors and seniors, documentation during device development, fostering economic incentives, and advanced internet/network security techniques. This exercise increased our confidence in our saturation criteria. Numerically, our taxonomy contains 265 categories (§3.4), so the ~3% potential increase is minimal. Taxonomically, inserting these topics and their corresponding recommendations into the taxonomy would not require a restructure. They would be located in levels 3 through 7, which is reasonably deep. Topics deeper in the taxonomy are more specific and, therefore, less concerning since they are not completely new ideas. Thus, we conclude that our saturation criteria are sound.

### 3.3 Taxonomy Validation

In this section, we describe our taxonomy validation methodology and results. Our goal is to measure the human bias in our qualitative categorization of recommendations [119].

**Methodology.** The validation was conducted by a fourth analyst who did not participate in the creation phase. They were given the taxonomy with all recommendations removed (*i.e.* removing the leaf nodes of the tree). We randomly sampled 10% of the recommendations (96/958 recommendations). The fourth analyst then remapped the sampled recommendations to the categories in the taxonomy. They could suggest new categories, but did not find this necessary. After the remapping, we compared the taxonomy to their categorizations.

We measured both exact agreement as well as the degree of disagreement. For exact agremeent, we calculated the agreement rate based on the proportion of matching categorizations.[5] For degree of disagreement, we used ConSim [124], which is used for ontology node similarity [49, 50, 53, 73]. ConSim measures the distance between two locations where a recommendation was placed (original coding vs. fourth analyst's coding). It rewards placements that are closer together (shorter path length) and further from the root node (less conceptual difference).

**Results.** The initial agreement rate between the taxonomy and the validator was 85.4% (82/96). After the discussion to reach a consensus, 6/14 mismatches went in favor of the original taxonomic categorization, resulting in an *adjusted* agreement rate of 91.6% (88/96) with 8 mismatches. The 8 mismatches had a near uniform distribution across ConSims of 0.0 and 0.9 (figure in supplementary materials), where a Consim of 0 indicates a worst-case mismatch (conceptually furthest in the taxonomy) and 1 indicates an exact match (conceptually closest in the taxonomy).

This indicates that our ConSim adjusted agreement rate is 95.5%, suggesting that the categorizations of the recommendations are accurate and the categories are well coded.

### 3.4 The Resulting Taxonomy

Similar to other taxonomies [94, 119], our taxonomy is also a tree (Figure 2). The taxonomy starts with a fixed root node (*i.e. IoT Security Recommendations*) whose two children correspond to the main distinction between security capabilities and secure processes (§3.2.1). The taxonomy's internal nodes represent our induced categories. Leaf nodes represent the actual recommendations. Here we give some preliminary analysis of the taxonomy: its categories, common and unique recommendations, contradictions, and target audiences.

**Categories.** At the first level, the taxonomy has two categories: (1) *Device Capabilities and Requirements* and (2) *Organizational Processes*. These are refined in subsequent levels. Under *Device Capabilities and Requirements*, there are 7 categories. Among these sub-categories, *Protecting Resources* contains the most recommendations (13% of total). These recommendations mostly focus on how the device can protect resources using software (*e.g.* authentication, access control, cryptography, and operating system capabilities) and hardware (*e.g.* cryptography, anti-tampering, and memory protection).

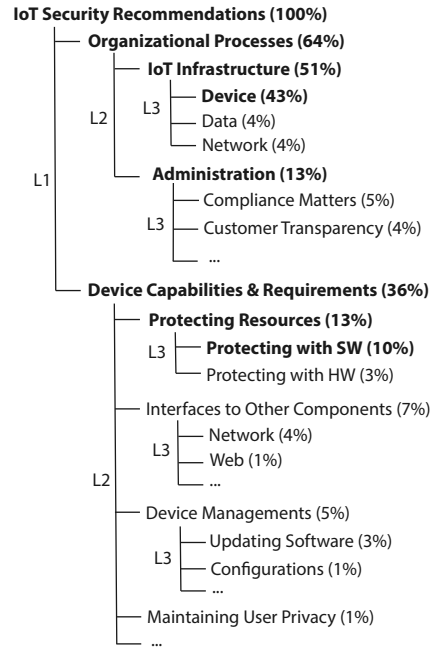

Fig. 2. Excerpt of the taxonomy. The tree contains sub-categories up to level 3 (% recommendation count in parentheses). Bolded categories contain 10% or more.

For *Organizational Processes*, there are 2 categories. *IoT Infrastructure* makes recommendations on how to securely handle each major component of the IoT infrastructure, which includes the device, data, and network. Recommendations for devices span its entire life cycle, with a subcategory for each stage: development, production, selection considerations, setup, management and operations, and disposal. The largest subcategory is management and operations (265 recommendations here vs 68 for production). *Administration* includes recommendations on processes that do not directly involve the IoT infrastructure but still might support it. In total, there are 265 categories (*i.e.* nodes in our taxonomy excluding recommendations), with 186 *leaf categories* (*i.e.* parent nodes to recommendations) and 79 *internal categories* (*i.e.* internal nodes, excluding leaf categories). The shallowest and deepest leaf categories are at levels 2 and 9, respectively. In contrast, the guidelines generally have, at best, very course-grained categories. For instance, UK-2018 has 13 categories in one level for their recommendations [116] and ENISA-2017 has 3 categories at level one and 24 total categories at level two [52].

---

[5]There is no benefit in calculating Cohen's kappa $\kappa$ because all agreement rates will be nearly equal to $\kappa$: the hypothetical probability of change agreement is negligible at $p_e = 1/265 = 0.004$, where 265 is the total number of categories in the taxonomy (§3.4).

**Recommendations.** We collected a total of 1,514 recommendations from the 25 guidelines listed in the supplementary materials. The taxonomy contains 958 unique recommendations, 728 of which are *exclusive* to the respective guideline and 229 of which had at least one duplicate. The most popular recommendation is to "*encrypt data in transit*". This result is not surprising since one of the core capabilities of IoT devices is to send data across networks.

Figure 3 shows both the total number of recommendations and the number of exclusive recommendations covered by the guidelines. Publisher-type-based analysis shows that the industrial non-profit sector contributed most of the recommendations in our corpus. Note that, the guideline IIC-2019 (Industry IoT Consortium) [69] has the most exclusive recommendations. It focuses on organizational processes, which other guidelines do not cover as much.

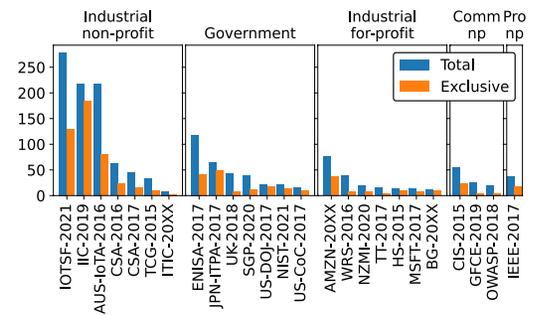

Fig. 3. Total and the exclusive number of recommendations per guideline. Exclusive recommendations mean that a guideline exclusively covers them.

**Any Contradictions?** Security advice from different sources can potentially create contradictions [98]. We consider a pair of recommendations contradictory if they are mutually exclusive, i.e., both can not be followed simultaneously. To check for contradictions, we examined the children of each leaf category in the taxonomy, as the leaf categories already group the related recommendations. For validation, two analysts reviewed all recommendations synchronously by traversing each taxonomy category. Any possible contradiction was discussed. We found only 3 pairs of contradicting recommendations (∼0.5% of all unique recommendations). The supplement lists all. We share one example here. The same guideline [43] recommended both to "*avoid words or phrases that include personal details in passwords*" [43], and to "*use a phrase that is long and personal for passwords*" [43]. Following both of the recommendations at the same time is impossible.

**Target Audience.** Based on the target audience as described in the guidelines, we organized the guidelines and recommendations into the following categories: Manufacturer (15 guidelines): designer, developer, or manufacturer of IoT devices and services; Organizational user (11): businesses and companies of any size that use IoT devices or services; Consumer user (5): household, family, or individual users of the IoT devices or services; Government user (4): users of IoT devices and services that belong to a part of a national government; and Government policymaker (1): any branch of a national government that has power to influence, modify, or create policies. Among the 958 unique recommendations, 65.5% target manufacturers and 52.5% target organization users. Only 5.7%, 3.2%, and 2.2% of the recommendations targeted consumer users, government policymakers, and government users, respectively. A recommendation may be target multiple audiences

## 4 THEME I: COMPREHENSIVENESS ANALYSIS

### 4.1 RQ1: Guideline Comprehensiveness

In this section, we discuss a method to measure guidelines' topic coverage and comprehensiveness by using a taxonomy. Next, we leverage our taxonomy to describe the topic coverage and comprehensiveness of the guidelines both individually and collectively. We report the comprehensiveness of individual guidelines by (1) breadth (number of topics in guideline), and (2) depth (number of recommendations per topic). We also investigate the topic coverage by publisher-type.

*4.1.1 Methodology.* First, we study each guideline's breadth and depth, which is commonly used to measure comprehensiveness in various domains [95, 110, 120]. To measure a guideline's breadth, we consider the number of leaf categories it contains, where more leaf categories indicate a larger breadth. To measure depth, other works assign assign a numerical score to each documents' coverage of a topic [48, 87]. For us, this is analogous to assigning a score equal to the number of recommendations in a leaf category for a guideline. Since we want to evaluate the depth across the entire guideline, we use the average number of recommendations per leaf category as our score, where a larger value indicates more depth.

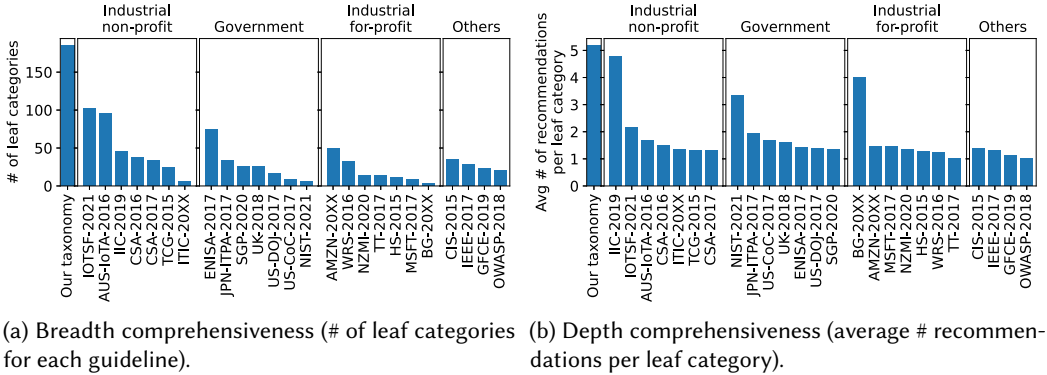

(a) Breadth comprehensiveness (# of leaf categories for each guideline).

(b) Depth comprehensiveness (average # recommendations per leaf category).

Fig. 4. Observe that IOTSF-2021, AUS-IoTA-2016, and ENISA-2017 cover the most diverse set of topics, as seen in Figure 4a. On the other hand, IIC-2019, NIST-2021, and BG-20XX have substantially more depth than the other guidelines, as seen in Figure 4b. This shows their emphasis on a certain set of categories in the taxonomy. Community non-profits and professional non-profits are combined into Others for clarity.

*4.1.2 Results.* For breadth, Figure 4a, shows that IoTSF-2021 [72], AUS-IoTA-2016 [71], and ENISA-2017 [52] are the only guidelines with 50 or more leaf categories, after which the number of leaf categories drops. BG-20XX [43] and NIST-2021 [89] have the first and second least leaf categories at 3 and 6, respectively. For depth, Figure 4b shows that IIC-2019 [69], BG-20XX [43], and NIST-2021 [89] are the only guidelines with more than three recommendations per category, with a drop after them.

**Overall Comprehensiveness.** Neither breadth nor depth individually captures the overall comprehensiveness of a guideline. To study the overall comprehensiveness, we plot the guidelines (Figure 5) with respect to both breadth and depth coverage. Here, we added our taxonomy for reference. Guidelines on the top right quadrant, though there aren't any, cover the most topics and go into the most depth for each category. These would be considered the most comprehensive. Next, the most comprehensive class of guidelines would fall into either the top left or bottom right quadrants. However, the guidelines in the top left quadrant cover a wide range of topics but may not cover them as in-depth as the guidelines in the bottom right quadrants. Similarly, the guidelines in the bottom right quadrant cover a few topics in depth but may suffer from low topic coverage. Guidelines with low comprehensiveness would fall into the lower left quadrant, covering the smallest range of topics in the shallowest depths. It can be seen that most guidelines fall into this quadrant. Interestingly, three guidelines from industrial non-profits and one from the industry for-profit organizations are significantly more comprehensive than others.

**Topic Coverage by Guidelines.** Our comprehensiveness analysis showed that a vast majority of the guidelines are significantly narrow and shallow. This motivated us to investigate them further from the perspective of broader categories. For this analysis, we use the categories from level two. We used the ratio of unique recommendations provided by a guideline in a certain category to the total number of unique recommendations in the same category as the metric for the measurement. A higher value indicates more comprehensive. In general, all the guidelines lack comprehensiveness for the individual broader categories too. Figure 6 shows that even the top 5 guidelines do not cover individual categories sufficiently. For instance, the guidelines' coverage ratios for *Organizational Processes* are mostly 0.1 and 0.2, with only IIC-2019 [69] reaching 0.3 and 0.4. This implies that the organizational process recommendations are spread evenly over many guidelines.

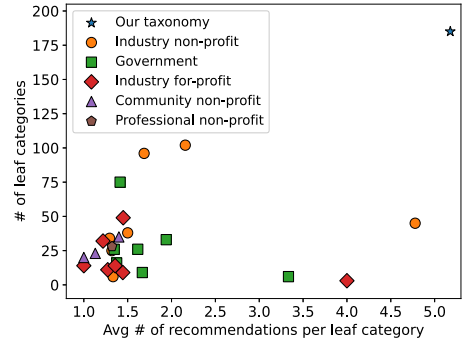

Fig. 5. Two-dimensional assessment of comprehensiveness for each guideline. Higher is better for both axes.

### 4.2 RQ2: Collective Topic Coverage

In this section, we report the topics covered by the guidelines collectively. Topics are represented by categories in the taxonomy.[6] We also report trends in topic coverage by publisher type and geographic region.

*4.2.1 Methodology.* We split the analysis into two parts. ❶ First, we look at the underlying statistical distribution of the topics as presented in the taxonomy. ❷ Then, we study the topic coverage based on publisher type and region, where *topic coverage* is defined as the ratio of the number of total recommendations from a guideline for a category and the number of all unique recommendations for said category. Topic coverage has a range between 0 and 1, inclusively. Publisher types are split into industrial for-profits, governments, industrial non-profits, and others [7]. Our publisher-type-based stratified sampling resulted in 16 guidelines from North America, 6 from Europe, and 3 from other parts of the world (Asia and Australia). Although incidental, this enabled us to look at region-based coverage. However, due to the small sample size of other regions, we only compare NA and EU. We chose categories from level two of the taxonomy for this analysis, as they provide a balance between specificity and generality.

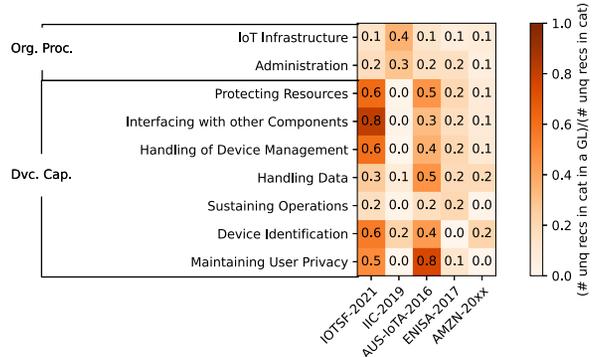

Fig. 6. The uniqueness of the guidelines, based on the proportion of unique recommendations. The level-two categories are shown for the 5 guidelines with the highest proportion.

---

[6]We use a different term from "categories" because "categories" are nodes of the taxonomy tree. The emphasis should be placed on the topics they represent, so we use "topics" instead.
[7]Community non-profit and professional non-profit were combined into others due to their small contribution.

*4.2.2 Results.* **Part ❶: Topic Distribution.** Figure 2 shows the topic distribution of the recommendations. At level one, 63% of all recommendations fall under the processes category, while only 37% of all recommendations fall under the device category. At level two, 50% of all recommendations are about processes regarding *IoT infrastructure*, while only 13% of all recommendations fall under *Administration*. None of the categories under the device capability exceeded 13%. Figure 2 also shows that only 1% of the recommendations are focused on privacy, 1% focus on safety, and 3% focus on the reliability of the device.

**Part ❷: Topic Coverage by Publisher-Type and Region.** Figure 7 shows the coverage of the guidelines based on the publisher type and region. When we look at the recommendations from different publisher types' perspectives, we see that most recommendations come from industrial non-profit organizations while the contributions from industrial for-profit seem minimal. Governments (7 guidelines) trail closely behind industrial non-profits. Guidelines from industry non-profits cover *Maintaining User Privacy* at 0.9, while 0.1 of them come from government publishers.

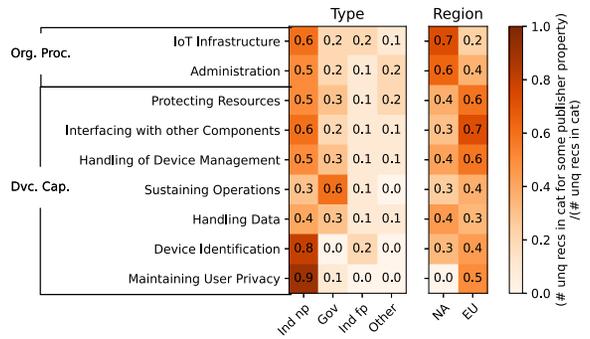

Fig. 7. Coverage across different publisher types and regions. Guidelines covered across publisher types are: Ind np (7), Gov (7), Ind fp (7), and Other (4). Comm np and pro np are combined into *Other due to the low sample size*. Guidelines covered across regions are: NA (16) and EU (6).

Our region-wise analysis shows that Europe (EU) and Australia (AUS) cover *Maintaining User Privacy* category at 0.5 and 0.6,[8] while North America (NA) and Asia do not cover them at all. We postulate that this may be because of two reasons. First, we searched exclusively for security guidelines and not privacy guidelines. They are related but separate domains. This is evident, as only 3/25 guidelines provided recommendations on privacy. Second, the EU and AUS have prominent data regulations in the GDPR 2016 [9] and Privacy Act 1988 [2], respectively. However, the United States, who is the major supplier of IoT guidelines in NA, lacks general privacy regulations at the federal level, although some exist at the state level (*e.g.* CCPA [8] and VCDPA [23]).

## 5 THEME II: USEFULNESS EVALUATION

In this section, we evaluate the usefulness of the recommendations and guidelines. We first assess individual recommendations by their *actionability* and *specificity* (§5.1). Then, we examine whether these recommendations could prevent CVEs and security failures (§5.2).

### 5.1 RQ3: To what extent are the recommendations actionable and specific?

*Actionability definition.* Not all recommendations are concrete enough to be actionable. For example, *"make privacy an integral part of the system"* [52] – it is not clear by the recommendation what action(s) to take to execute it. We consider such goal-oriented recommendations as *unactionable*. On the other hand, a recommendation is *actionable* if contains concrete actions. For instance, the recommendation that *"authentication credentials shall be salted, hashed and/or encrypted"* [52] is actionable as it prescribes specific actions.

---

[8]Note that the summation of percentage coverage might exceed 1, because there can be multiple guidelines making duplicate recommendations. For instance, we have category *Handling Data* with % coverage of 0.4, 0.3, and 0.5 for guidelines from NA, EU, and Others regions.

*Specificity definition.* Like privacy policies for smart home devices [80], recommendations can vary widely in semantic specificity. Here, we study the specificity of the recommendations in terms of their ability to prevent specific threats, vulnerabilities, and/or attack surfaces, if executed correctly. A recommendation using broad terms, by definition, will not be able to prevent something specific. For instance, the recommendation that *"the web user interface shall follow good practice guidelines"* [72] does not specify which specific good practices for web user interfaces to follow, so it is unclear which security issues it can prevent, if any. On the other hand, the recommendation that *"authentication credentials shall be salted, hashed and/or encrypted"* [52] can prevent the specific issue of the device storing passwords in plaintext, *i.e.* unsalted and unhashed.

*5.1.1 Methodology.* Because we need to review all 958 recommendations, for actionability, we simplify the labeling process by asking only one question: *After reading the recommendation, will the reader have a reasonably high chance of knowing what action(s) he/she should take to execute the recommendation?* If yes, then the recommendation is actionable. To determine whether a recommendation is specific or not, we ask the question: *Can the recommendation prevent specific threats, vulnerabilities, and/or attack surfaces from the device, if implemented correctly?* If yes, then it is specific, and non-specific otherwise. To label all the recommendations, we used an iterative coding process [113], where two authors meet synchronously to review the generated labels and discuss any disagreements until mutual agreement.

*5.1.2 Results.* We found 836/958 (87.2%) of the recommendations in our corpus to be actionable. We found 371/958 (38.7%) of the recommendations to be specific. Figure 8 shows the distribution of recommendations and their actionability and specificity labels in the level 2 category. There are 373 (38.9%) actionable and specific recommendations, 463 (48.3%) actionable but not specific recommendations, and 122 (12.7%) neither actionable nor specific recommendations. Most device capability recommendations are actionable and specific: *e.g.* "*Encryptions keys must be truly randomly internally generated*" [72]. Actionable but not specific recommendations are often about supportive processes that help with security indirectly: *e.g.* "*The contract with the vendor must include an audit clause for one security assessment per year at a minimum*" [47]. Interestingly, there are 0 specific but not actionable recommendations. This makes intuitive sense because if a recommendation is not actionable, it will not be able to prevent any security issues. In addition, unactionable recommendations are often goal-oriented, *e.g.* the recommendation "*minimize exposed attack surfaces*" [61] does not specify which attack surfaces.

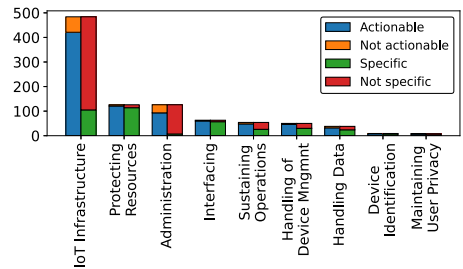

Fig. 8. Number of actionable, not actionable, specific, and not specific recommendations at each level 2 category. We see that, infrastructure and administrative recommendations largely do not address specific security issues (*i.e.* not specific).

*5.1.3 Measuring Bias with a User Study.* **Method.** To measure the human bias in our *actionability* and *specificity* labels, we conducted an IRB-approved user study [51]. We randomly selected 100 recommendations to be labeled by independent analysts. For the analyst population, we preferred a different population from the research team to reduce selection bias and expose any conformity bias among the researchers. In addition, we preferred the analyst population to represent a lower bound in terms of knowledge and experience with respect to the entire spectrum of software engineers. As fourth-year university students are often used in experiments as proxies for junior software engineers [33, 54, 64, 109], we chose them (who took a senior-level computer security course) as our study population. Using the same questions from §5.1.1, we designed a questionnaire. To identify any problems with the questionnaire, we conducted a pilot study [51], before the actual study. The final survey is in the supplemental materials.

In total, we invited 15 students [9] to review 20 recommendations such that each of the 100 recommendations is reviewed by exactly three students. The final label is determined by the majority voting. However, we discarded votes if the student did not sufficiently understand the recommendation. In the end, we had 43 and 21 labels for actionability and specificity, respectively.

**Results.** The agreement rates between the author and students for actionability and specificity are 83.7% (36/43) and 76.2% (16/21). This results in a Cohen's kappa of 0.67 and 0.52 (average 0.62) for actionability and specificity, respectively [83], which Cohen suggests can be interpreted as substantial and moderate agreement.

## 5.2 RQ4: To what extent could the guidelines prevent real-world vulnerabilities?

Here, we define recommendations as "useful" if they could mitigate real-world security problems. Because the majority of the guidelines target manufacturers and organization users (65.6% and 52.5%, respectively), our analysis concerns only manufacturers and organization users. Specifically, we pose the following research question: *To what extent could the recommendations have prevented developers from making design mistakes leading to vulnerabilities and real-world security failures?*

*5.2.1 Methodology Overview.* We answer this with a retrospective study by analyzing past events [81]. Retrospective studies is a form of observational study used in many other research fields (*e.g.* healthcare, medicine, psychology) when conducting a prospective study is deemed infeasible or unnecessary [81]. For this work, the past events are security vulnerabilities recorded in Common Vulnerabilities and Exposures (CVEs) and real-world failures reported in the news media.[10] For each CVE or failure, we find recommendations from our taxonomy, that could have prevented them. Results are reported in §5.2.5 and §5.2.6, respectively. For better generalization, we categorized the CVEs and failures into the following seven application domains, expanding the categories from [30] based on our observations: (1) consumer product, (2) commercial product, (3) industrial product, (4) medical product, (5) automotive product, (6) critical infrastructure, and (7) SW/HW component. These categorizations were done by one author and validated by another author; there were no disagreements.

*5.2.2 CVE Collection Methodology.* CVE analysis requires a set of IoT-related keywords to search for them.[11] To gather the set of keywords, we collected survey papers published from 2017-2022 in the journals: (1) *IoT Journal*, (2) *ACM Computing Surveys*, (3) *Transactions on Software Engineering*, and (4) *Transactions on Dependable and Secure Computing*. We included any paper whose title matched "survey", "security | privacy" and "IoT | internet of things", resulting in 4 papers. We read these papers and extracted any terms related to (1) IoT vulnerabilities, (2) IoT devices or products, and (3) IoT software packages and libraries, resulting in 36 keywords. Using these keywords, we searched for CVEs from 2011-2022 that match ≥ 1 keyword in their description. Next, we extracted the names of companies from these CVEs and determined the names of their IoT-related products. Finally, we use pairs of company names plus product lines to serve as the keywords to search in the CVE database, filtering out any non-IoT CVEs. The resulting list contained 2500 CVEs from 61 companies [12] However, the distribution of CVEs is heavily skewed by large vendors like Apple and Qualcomm. Therefore, we used stratified random sampling, where each stratum is a company, to randomly choose up to 4 CVEs per company. Our final list contained 158 IoT CVEs.

---

[9]The students were compensated $20 for their participation.
[10]Since no scientific corpus of public IoT failures exists, we relied on public news reports to build one for this purpose.
[11]A trial run using keywords from IoT literature review papers did not yield a rich corpus of IoT CVEs.
[12]Please see Figure 3 in Appendix.pdf [24].

*CVE Preparation.* The goal of CVE analysis is to study if IoT-related CVEs could have been prevented by following the recommendations in our taxonomy. However, CVEs originating from coding errors like buffer overflows can only be minimized, not completely avoided, even if following recommendations like conducting static analysis [72]. On the other hand, CVEs caused by design flaws, *e.g.* a device having "*hard-coded system passwords that provide shell access*" [16] can be prevented if a corresponding recommendation exists (§5.2.5). Thus, we filtered them out by using the cognitive distinction made between two kinds of errors: *implementation slips* (right plan, but made an unintentional coding error) vs. *design mistakes* (wrong plan, leading to a design flaw) [93]. To this end, one author first labeled each CVE independently (slip or mistake) and then, for validation, discussed the labels with another author until agreement, filtering out 61 slips (Table 1). In total, we collected 97 design-mistake CVEs; these are the CVEs to which recommendations will be matched.

Table 1. Implementation slips.

| Type | Count |
| --- | --- |
| Buffer overflow | 18 (30%) |
| Another memory issue | 8 (13%) |
| Validation logic | 7 (11%) |
| Not enough info | 19 (31%) |
| Other | 9 (15%) |
| Total | 61 |

*5.2.3 News Collection Methodology.* Following the methods of an IoT failure study [29], we compiled IoT security failures reported by reputable news sources: *WIRED* and *The New York Times*. For reproducibility, we used Google News to search both sources. We used a condensed list of search terms from phrases used for the CVE search. [13] Specifically, we conducted a pilot search to eliminate phrases that did not yield results and condensed common terms across multiple phrases. We also added general terms such as "autonomous", "cyber-physical system", and "cyber security fail" to collect a wider set. We limited the search from January 2015 to October 2021 [14]. The results were filtered for articles describing a security failure, first by title and then by content. The article with greater detail was used, when multiple articles described the same failure. Sources referenced in the articles were also reviewed for supplementary information. Our search criteria yielded a total of 1400+ news article results, out of which we identified 28 IoT security failure articles. However, we discarded 11 of the them due to insufficient information. Of the remaining 17 news articles, 4 were reported by *The New York Times* and 13 by *WIRED*. Across the 17 articles, there were 33 sources of failure and 29 repair recommendations; most articles listed multiple sources of failure and multiple repair recommendations.

*5.2.4 Recommendations Matching.* We do the following to match recommendations to a given CVE or news article. First, we choose the category at each level of the taxonomy that is most related to it, traversing down until we reach recommendations (leaf nodes). This step helps us find the set of most relevant recommendations. If there is no such category, then there is no matching recommendation. Note that we matched only device recommendations to the *design-mistake-CVEs*, while both device and process recommendations are matched to news. Then, if we do find the optimal category, we look through each recommendation and only consider recommendations labeled as *specific* (§5.1) and *actionable* (§5.1). Note that some recommendations reported by the journalists are for handling and post-analysis of failures, not preventing them (*e.g.* "*consider introducing a mechanism for communicating to others when the system is in the degenerate operation mode*" [74]). To retain this information, we also allow non-specific recommendations for news stories. To avoid inconsistencies and bias, two different authors independently performed the identifications and mapping for CVEs and news stories, and two others validated them (senior security researchers). The validation was done by checking whether the selected recommendations could prevent the CVEs or failures.

---

[13]Search Terms: "iot, cyber-physical system, cyber security fail, autonomous, smart, Fitbit, embedded device, robot, wearable, industrial control, router, sensors".
[14]Google News Search Syntax: "[SearchTerm] site:[SourceWebsite] after:[StartDate: Year-Month-Day] before:[EndDate: Year-Month-Day]."

Additional recommendations that might have addressed the CVEs/failures were also identified. Finally, the authors discussed any disagreements until reaching an agreement. Since the validators did not independently create new mappings but worked with the original ones, calculating interrater reliability would not add meaningful insights.

*5.2.5 CVE Analysis Results.* **Collective Analysis.** We define *% coverage* as the percentage of design issue CVEs for which a guideline provided one or more recommendations. We were able to pair one or more recommendations from our taxonomy to 77 of the 97 design issue CVEs, for a coverage of 79%. The potential impact of the taxonomy on the CVEs are outlined in Figure 9. We found 44 recommendations from 15 categories. Below are two case studies.

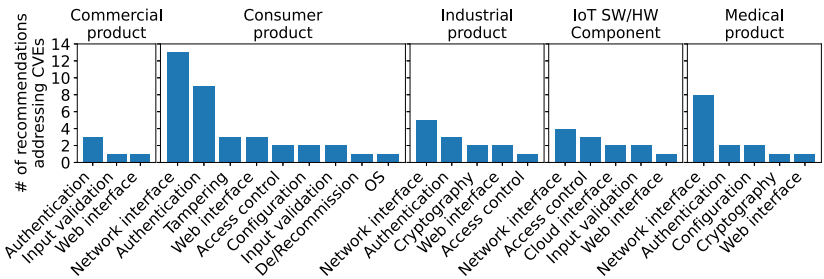

Fig. 9. Number of recommendations addressing CVEs from different application domains. There are no process recommendations since we only match device recommendations to CVEs. Here, network interface-related CVEs have the most matching recommendations, reflecting IoT devices' core functionality of networking.

**Case study (covered):** Consider CVE-2021-33218 [16]: "*...There are Hard-coded System Passwords that provide shell access*". This CVE could have been mitigated by the following recommendation: "*Security parameters and passwords should not be hard-coded into source code or stored in a local file*" [72].

**Case study (not covered):** Consider CVE-2021-27943 [17]: "*The pairing procedure ... is vulnerable to a brute-force attack (against only 10000 possibilities), allowing a threat actor to forcefully pair the device, leading to remote control of the TV settings and configurations*". While there were recommendations suggesting to prevent brute-force account logins, there were none for brute-force device pairing. Although not all IoT devices may necessarily require such pairing capabilities, it is indeed one way of connecting to networks, which is a capability by definition all IoT devices have.

**Individual Analysis.** In Figure 10a, it can be seen that most (24/25) guidelines have CVE coverage of less than 40%. In fact, only IOTSF-2021 [72] achieved CVE coverage of over 50%, with 62% coverage. Although IIC-2019 [69] has the second most total recommendations (200+), its emphasis on organizational processes and not device capabilities puts it in 9th place with 14% coverage. Meanwhile, 7/25 guidelines had 0% coverage. For instance, NIST-2021 [89] provides recommendations for "establishing IoT device cybersecurity requirements", which focuses on organizational processes for integrating IoT devices to current infrastructure. None of these 7 guidelines targeted manufacturers or had any recommendations under the device features category.

*5.2.6 News Analysis Results.* **Collective Analysis.** We were able to match one or more recommendations from our taxonomy to all 17 news stories for a coverage of 100%, as seen in Figure 10b. The potential impact of the taxonomy on the failures are outlined in Figure 11. We found a total of 41 recommendations from 13 categories. The leading mitigation categories were network segmentation (6/17 news stories), network security (6/17), authentication (4/17), and access control.

**Case study:** To illustrate our results, we present an example using news article 3, reported by WIRED in 2019 [12]. This article reported security vulnerabilities in the GPS tracking apps, iTrack, and ProTrack. These apps are used to track fleets equipped with GPS trackers, such as GT06N by Concox. The article reports that these apps were exploited to gain unauthorized access to thousands of vehicles. The exploit exposed access to safety-critical functions of the vehicles, such as the remote engine power toggle. This article described two sources of failure: *i)* A default password was enabled across all accounts, and *ii)* The GPS tracking device access was not isolated from a safety-critical function (engine on/off). We found 8 recommendations that could have mitigated these failure causes, coming from the *Authentication* and *Network Segmentation* categories. Two of these recommendations were, from *Authentication*: "*Each device should have a unique default password*"[68]; and from *Network Segmentation*: "*[Split] network elements into separate components to help isolate security breaches and minimize overall risk*"[52].

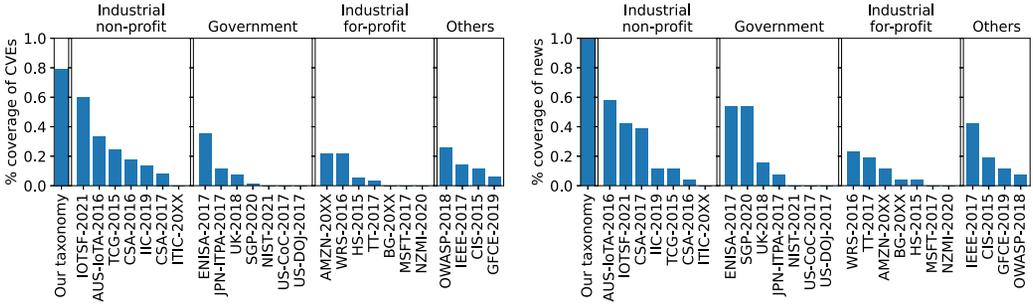

(a) Coverage of CVEs for each guideline.  (b) Coverage of news for each guideline.

Fig. 10. Coverage of CVEs and Failures for each guideline. Here, the number of CVEs and failures is normalized. Community non-profits and professional non-profits are combined into Others for clarity.

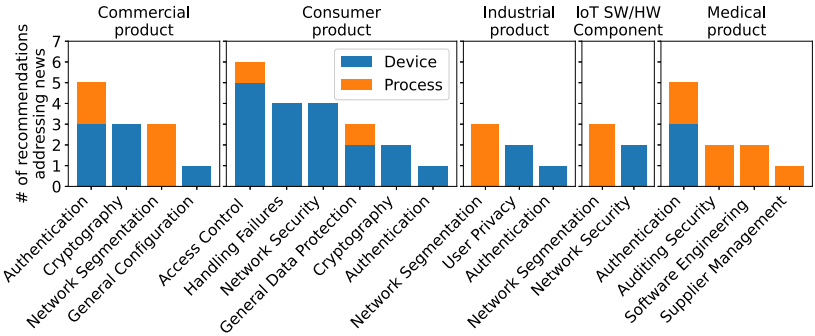

Fig. 11. This figure shows recommendations addressing IoT security failures reported in the news, where there are both device and process recommendations. A notable observation is that Authentication appears in 4/5 clusters and Network Segmentation appears in 3/5 clusters, suggesting that these issues are widespread.

**Individual Analysis.** Figure 10b reports the coverage of failures for each guideline. 19/25 guidelines had coverage of less than 40%, and 6/25 guidelines had 0% coverage. The rest of the guidelines only had 40% to 60% coverage. The guideline with the most coverage was AUS-IoTA-2016 [71], followed by ENISA-2017 [52] and SGP-2020 [107]. For instance, US-CoC-2017 [89] provides very few recommendations, limiting its application to real-world failures.

## 6 DISCUSSION

### 6.1 Statement of Positionality

Despite our efforts to reduce bias throughout the work, they may still persist. Therefore, for transparency, we note our positionality here. The research team consists of engineers and researchers. Two of the four authors have over 3 years of industry experience as software engineers, in addition to research experience in security and software engineering. One of these two authors also has extensive experience in qualitative research methods. Overall, their comprehensive experiences allow them to advise the other two authors (graduate students) in all tasks and experiments conducted in this work. However, our perspectives and experiences may shape our interpretation of the guidelines. For example, our team lacks legal and organizational management expertise, which may have influenced our interpretation of the guidelines.

## 6.2 Comments on Guidelines

**Inadequacy of the Guidelines.** Our comprehensiveness analysis showed weaknesses in most of the IoT guidelines (§4). Only four of the guidelines are reasonably comprehensive: three from industrial non-profits and one from an industrial for-profit organization. Of these four, one covers depth while the others cover breadth. Even after combining all guidelines into a unified taxonomy, our usefulness analysis (§5.2) shows that it does not mitigate 21% of the CVEs in our corpus. On closer investigation, we suggest that IoT technologies are too diverse to be captured by general guidelines with a broad definition of IoT, which is even acknowledged by a guideline we reviewed [115]. For example, six guidelines define IoT as anything that connects to the Internet or a network and interacts with the physical world and data [25, 52, 89, 107, 117, 123]. Recall our comprehensiveness analysis on broader categories (§4.1.2) — we found that individual guidelines were not comprehensive even for general topics such as passwords. We, therefore, suggest that guideline publishers develop more domain-specific guidelines with better comprehensiveness. Defining security requires context; security guidelines must define a context, specific and concrete enough to be actionable.

**Security vs Cost Tradeoff.** Security engineering is not free, thus, following security recommendations is hindered by its associated costs. Our unified corpus contains 958 unique recommendations — this is a lot for an organization to track. We suggest that guideline providers incorporate a priority score for their recommendations based on their importance. Only IOTSF-2021 [72] and IIC-2019 [69] provided metrics for which recommendations to follow in order to meet certain security statuses, thereby implying their importance. This would facilitate a better cost-vs-security tradeoff for companies with constrained compliance budgets [35, 62].

## 6.3 Implications

**Policymaking.** Current legislation on the security of IoT systems appear to adopt only common recommendations from guidelines. Current IoT security policies are limited to discussions of "reasonable security features" that are limited to how passwords are set [37]. For example, the first IoT-specific security law passed in 2018 in the United States, California Senate Bill No. 327: "Information privacy: connected devices" primarily contains two common security recommendations as laws that require devices to have (1) unique preprogrammed passwords and (2) a feature to require users to change passwords during set-up [7]. The guidelines from which policymakers review and adopt recommendations will determine the comprehensiveness of legislation. Given that only 4/25 of the guidelines we studied were reasonably comprehensive and that 19/25 guidelines had coverage less than 40% to prevent newsworthy failures, the current reality seems to be concerning.

**Practitioners.** Given that most guidelines are not comprehensive, practitioners should consult multiple guidelines depending on their security priority. Generally, the most comprehensive guidelines are from government and industry non-profits, so practitioners can prioritize these over other publishers. In addition, comprehensiveness is correlated with the number of recommendations and pages. We recommend beginner practitioners consult longer guidelines and consider which security capabilities or processes would be most impactful to them, as no guidelines contain priority scores.

## 6.4 Threats to Validity

*6.4.1 Construct Validity.* In §3, we operationalized the constructs of "general IoT security guidelines" and "recommendations". Our definitions are generally consistent with prior work [34, 36]. We did broaden the definition of "guideline" to results from search engines, rather than relying on the definition by the UK government [116]. Since we obtained guidelines from many prominent providers including some omitted by prior work, we think this was a sound decision.

In §4.1, we compared the breadth and depth of a given guideline across all the guidelines in our corpus. This is natural because our corpus is general-applicable to all fields. However, we recognize that general guidelines might not cover all the aspects of IoT security, especially for different fields of applications. This is indeed the main purpose of designing metrics like our proposed measures of breadth and depth coverage so that practitioners can use this to make informed decisions on whether a given guideline is sufficient for their purpose.

In Theme 2, we operationalized the construct of "usefulness" using notions of "actionability" and "specificity". There is disagreement in the literature on appropriate definitions. For example, Redmiles *et al.* measured perceived actionability with sub-metrics including time and difficulty [98]. Barrera *et al.* used a decision tree, *e.g.* assessing language quality and resource needs [34]. We believe these approaches do not directly apply in our study. For example, the target audience for [98] is the general population, rather than engineers. Similarly, our recommendation extraction method excluded recommendations with low language quality by design, so the corresponding questions in [34] are irrelevant. Given the nature of the recommendations in our taxonomy, we think our qualitative approach is more reasonable.

*6.4.2 Internal Validity.* Our analysis depends on the taxonomy, which was obtained manually. To increase the trustworthiness of the taxonomy, we followed standard practices in qualitative methods, including those from taxonomy development. To mitigate bias, we conducted a pilot study and several rounds of multi-person scrutiny. We also validated the taxonomy with an independent analyst, achieving 96 % agreement (adjusted for conceptual similarity). An alternative to manual taxonomy construction is the use of large language models, which may induce the biases of the model [59, 65]. Finally, while news reports may contain dramatization and selectively cover events [88], news has been used as a primary data source for research in computing [29], civil engineering [128], and public health [57].

*6.4.3 External Validity.* We defined the population of general IoT security guidelines via web search engines, on which we used stratified sampling. Several threats result. (1) The selected guidelines are all free — we omitted paid guidelines such as those by ISO and IEC. (2) We only considered guidelines written in English. (3) We considered guidelines individually rather than by publishing organization — some organizations publish recommendations across multiple guidelines, so our results hold for individual guidelines but do not generalize to the corresponding publisher. (4) We studied only general IoT security guidelines and not those for specific domains — our taxonomy may miss topics only covered by guidelines for specific domains. However, despite these concernces, many CVEs and all of the news failures could be mitigated using these guidelines (in aggregate).

Furthermore, while the recommendations can be evaluated by conducting a retrospective study with IoT CVEs and failures in the news, this may not be an accurate reflection of reality. For instance, implementation mistakes may occur even after reading the recommendations, since they only aim to provide knowledge. Finally, even though we employed a strict saturation criteria to determine the stopping point for guideline reviewing, none can guarantee the knowledge coverage of all guidelines. However, our validation of the stopping point suggests that our analysis misses minimal new insights from the remaining guidelines.

## 7 RELATED WORK

In §2, we described prior work describing IoT engineering practices and security vulnerabilities. Here, we present two more distant strands of work.

**Security Policy Automation.** Automation can help reduce the cost of complying with IoT security guidelines. Natural language processing shows early promise for automatic analysis of guidelines [63] to extract security recommendations. Once recommendations are extracted, some aspects might be automatically checked [56, 75, 96, 122, 126]. Notably, Ferraiuolo *et al.* demonstrated an automated approach for automatic software supply chain integrity checks [56].

**Studies on Software Engineering Guidelines.** Prior works have also studied software security guidelines for digital systems [111], privacy guidelines for modern databases [105], mobile financial service applications [42], children's Android applications [99], and the C programming language [39, 67]. One work studied how security advice are produced through a series of interviews [92]. Research has also been done in privacy policies for smart home devices [80].

## 8 CONCLUSION AND FUTURE WORK

Many IoT security guidelines have been proposed to improve the security of IoT systems. Our goal was to assess the coverage and usefulness of these guidelines, both collectively and individually. We developed a taxonomy of recommendations extracted from 25 IoT security guidelines. *Our findings reveal that even the most comprehensive guidelines do not cover security topics widely or deeply. Our usefulness evaluation shows that most guidelines alone will not suffice to prevent all IoT-related security failures.* We found that, collectively, the recommendations address 77 out of 97 design-related CVEs, as well as all 17 security failures in the news. However, only one out of 25 guidelines covered over 40% of the CVEs, and just four out of 25 guidelines covered over 40% of the news stories. The **implication** for developers is that they should refer to multiple guidelines for in-depth defense, as a single one may not suffice. Publishers should, based on the goal, consciously choose what breadth and depth a guideline should cover and explicitly state it in the guideline.

However, there is minimal information about whether and how guidelines are used. To this end, we suggest three directions for **future work**. First, further qualitative analysis on the guidelines is needed. For example, measuring the correctness of guidelines — including system and threat models, currently omitted from guidelines — would be valuable. Second, little is known about the use and utility of such guidelines in practice. Empirical studies would benefit IoT manufacturers, users, and guideline authors. Another potential future work is conducting prospective observational studies where developers implement certain parts of an IoT device, or IT specialists implementing certain processes. Our results show that such studies should include multiple guidelines for comprehensiveness, which can be informed by this study.

### DATA AVAILABILITY

A replication package was uploaded as supplemental material [24] , containing: (1) a list of studied guidelines and extracted recommendations; (2) the full taxonomy with a tool to view it; (3) validation data from the CVE and news article analysis; and (4) anonymized results from our study of novices.

### ACKNOWLEDGMENTS

We thank the anonymous reviewers for their feedback. We thank Dr. David Lo for his help throughout the submission process. This work was supported by the IT4IR TRIF program from the University of Arizona's Office of Research, Innovation & Impact (RII).